\documentclass[aps,prl,twocolumn,superscriptaddress,showpacs,showkeys]{revtex4}
\usepackage{bm,amsmath,graphicx,dcolumn}
\begin{document}

\title{An X-ray vision of cataract}
\author{Andrea Antunes}
\affiliation{Instituto de F\'{\i}sica, Universidade de S\~ao Paulo, CP 66318, 05315-970 S\~aoPaulo, SP, Brazil}
\email{antunes@if.usp.br}
\author{Ang\'elica M.V. Safatle}
\affiliation{Faculdade de Medicina Veteri\'aria, Universidade de S\~{a}o Paulo, 05508-900 S\~aoPaulo, SP, Brazil}
\author{Paulo S.M. Barros}
\affiliation{Faculdade de Medicina Veteri\'aria, Universidade de S\~{a}o Paulo, 05508-900 S\~aoPaulo, SP, Brazil}
\author{S\'ergio L. Morelh\~ao}
\email{morelhao@if.usp.br}
\affiliation{Instituto de F\'{\i}sica, Universidade de S\~ao Paulo, CP 66318, 05315-970 S\~aoPaulo, SP, Brazil}

\date{\today}
\begin{abstract}
This work reports the exploitation of diffraction enhanced X-ray imaging (DEI) for studying cataract in addition to the finding of heavy scattering centers of light, probably Ca-rich precipitates in cataractous lenses. DEI selectively probes diffuse-scattering, refraction, and absorption properties of features in the lenses. Fiber cell compaction areas and dilute distribution of precipitates are identifiable, as well as highly absorbing aggregations providing contrast even in the conventional radiography setup. This finding opens new opportunities for clinical diagnosis, for understanding the causes of cataract, and in developing medicines for this disease.
\end{abstract}



\maketitle

\section{Introduction}

Cataract, or eye lens opacity, is an important ophthalmic disease \cite{jaff1992,brow1996} and the leading cause of eye surgeries in the world, affecting the life quality of millions of people. Intraocular lens implantation is a successful treatment, minimally invasive with quick rehabilitation, low complication rate, and affordable for most populations. However, it is still necessary to understand the direct causes of this disease at a molecular level in order to prevent, postpone, or correct the vision loss at early stages of cataract.

Almost all of the transparency loss is due to light scattering and, consequently, the subject of most works done in recent years is to identify the mechanisms by which the biochemical or cellular changes lead to the visible light scattering that clouds the lenses \cite{ghou1996, dill1999, tayl1999, trus2003, tang2003, bass2003}. Such scattering requires abrupt density variations, or refractive index fluctuations, in a length scale comparable to the wavelength of the visible light \cite{gill2001}. Fiber cell compaction in age-related cataract \cite{ghou2001,free2003} produces density variations over much larger scales to explain alone the diffuse light scattering observed in the catarogenesis.

Morphological studies are intimately related to the availability of characterization techniques. Since visible light microscopy in the later 19th century \cite{rabl1900} to transmission and scanning electron microscopy of our times \cite{jong1998,free2003,kusz2004}, the lens structure were extensively studied. Acoustic and ultrasonic properties \cite{kort1994,taba2000} and Raman signature \cite{erck2001,guti2004,antu2005} of healthy and cataractous crystalline lenses have also been reported. 

Sophisticated optical imaging techniques \cite{mast1998,dica1999,kuro2002} are important tools for {\em in vivo} clinical diagnosis of cataract, and to establish statistical correlations between the patient's medical history and the degrees and types (nuclear/cortical) of cataract.

\begin{figure}
\includegraphics[width=1.6in]{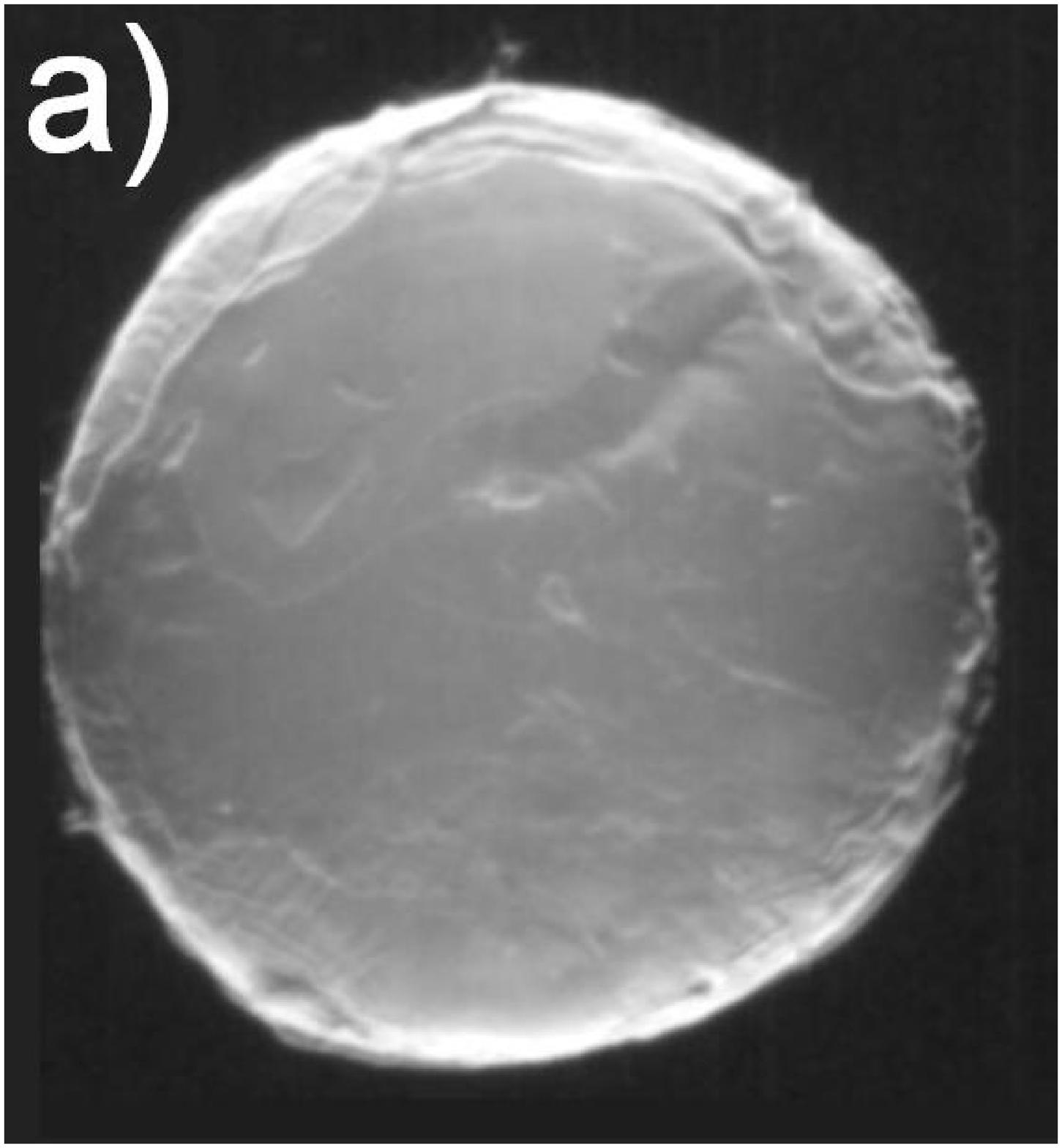}
\includegraphics[width=1.6in]{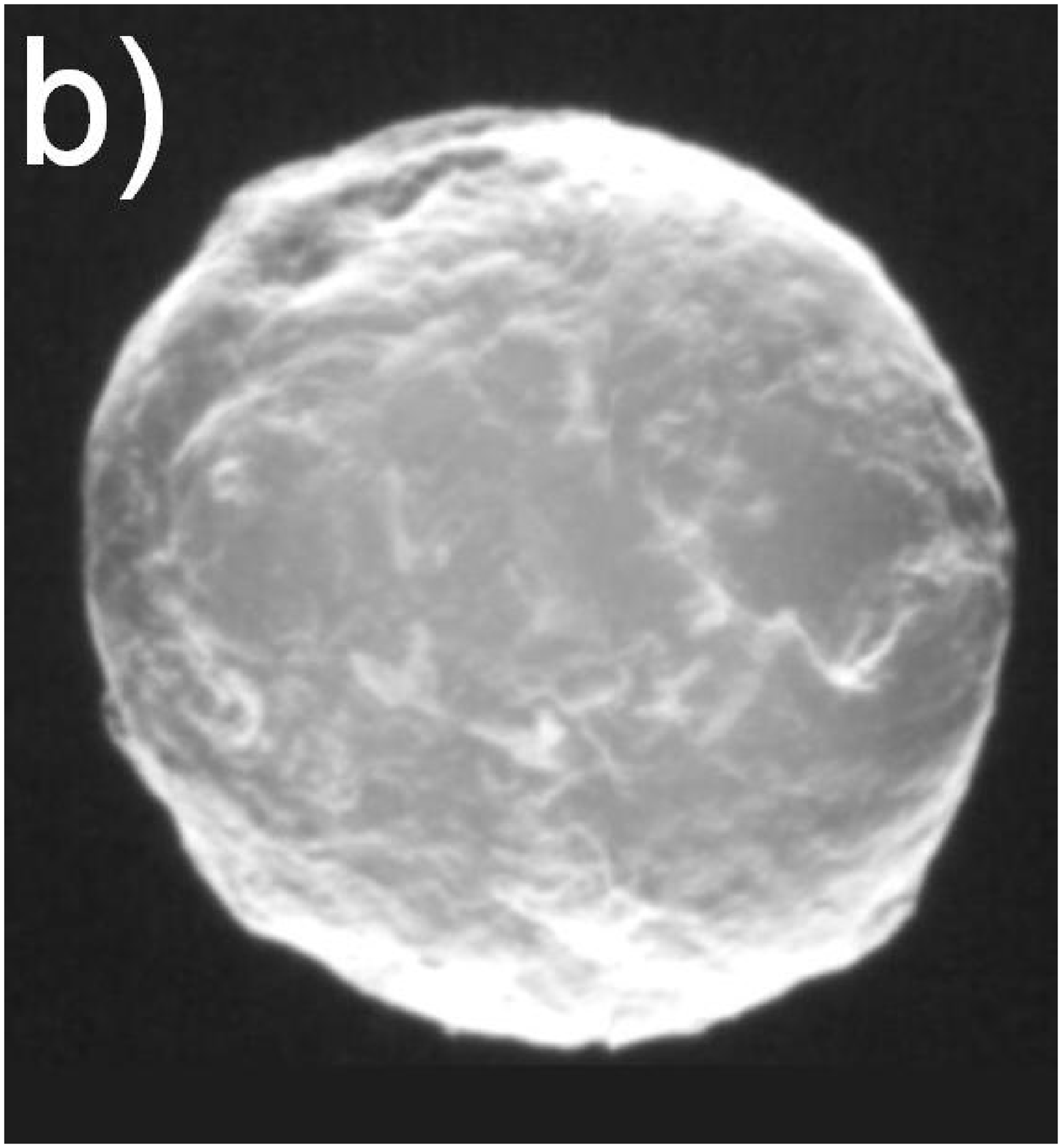}\\
\includegraphics[width=1.6in]{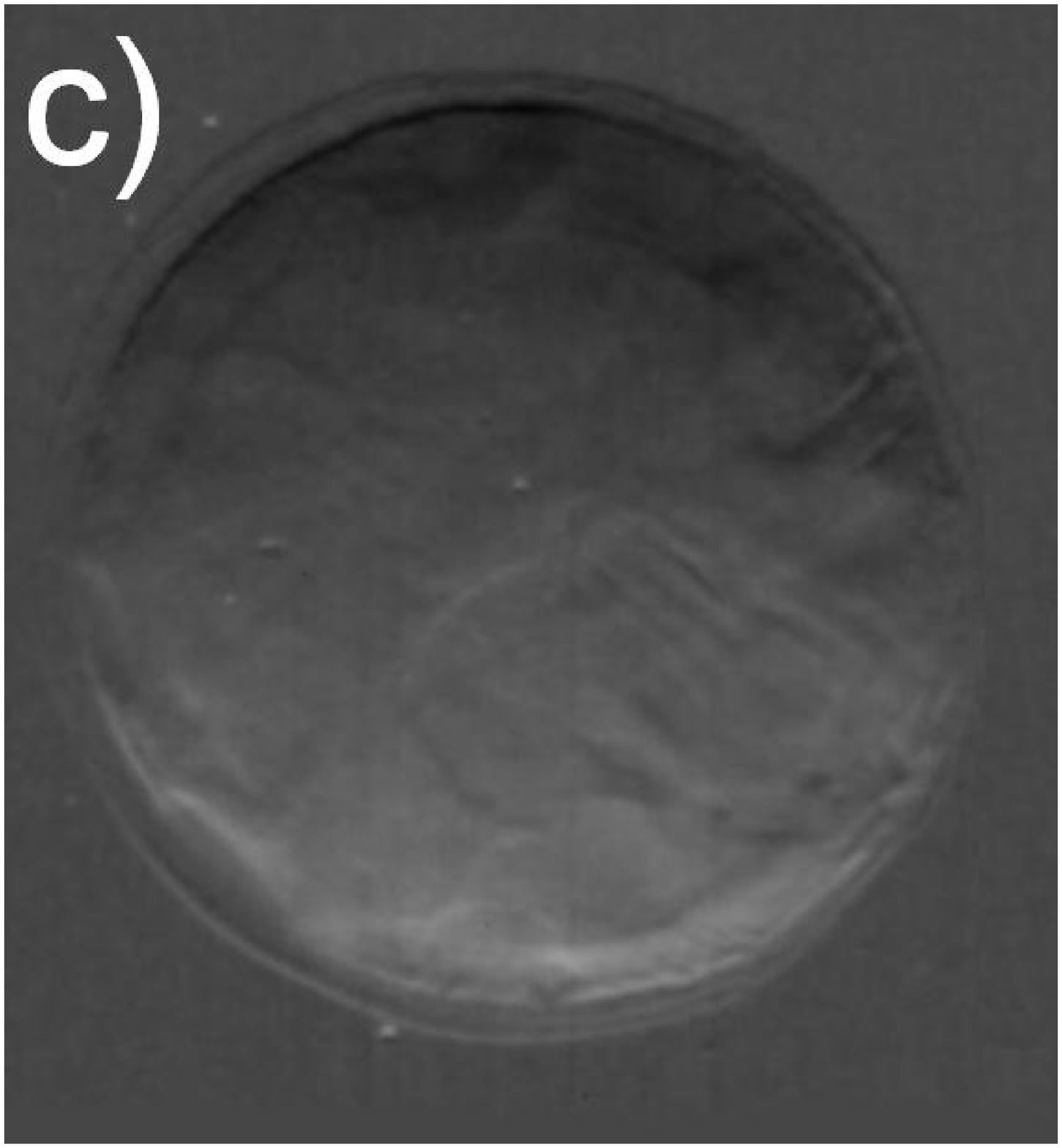}
\includegraphics[width=1.6in]{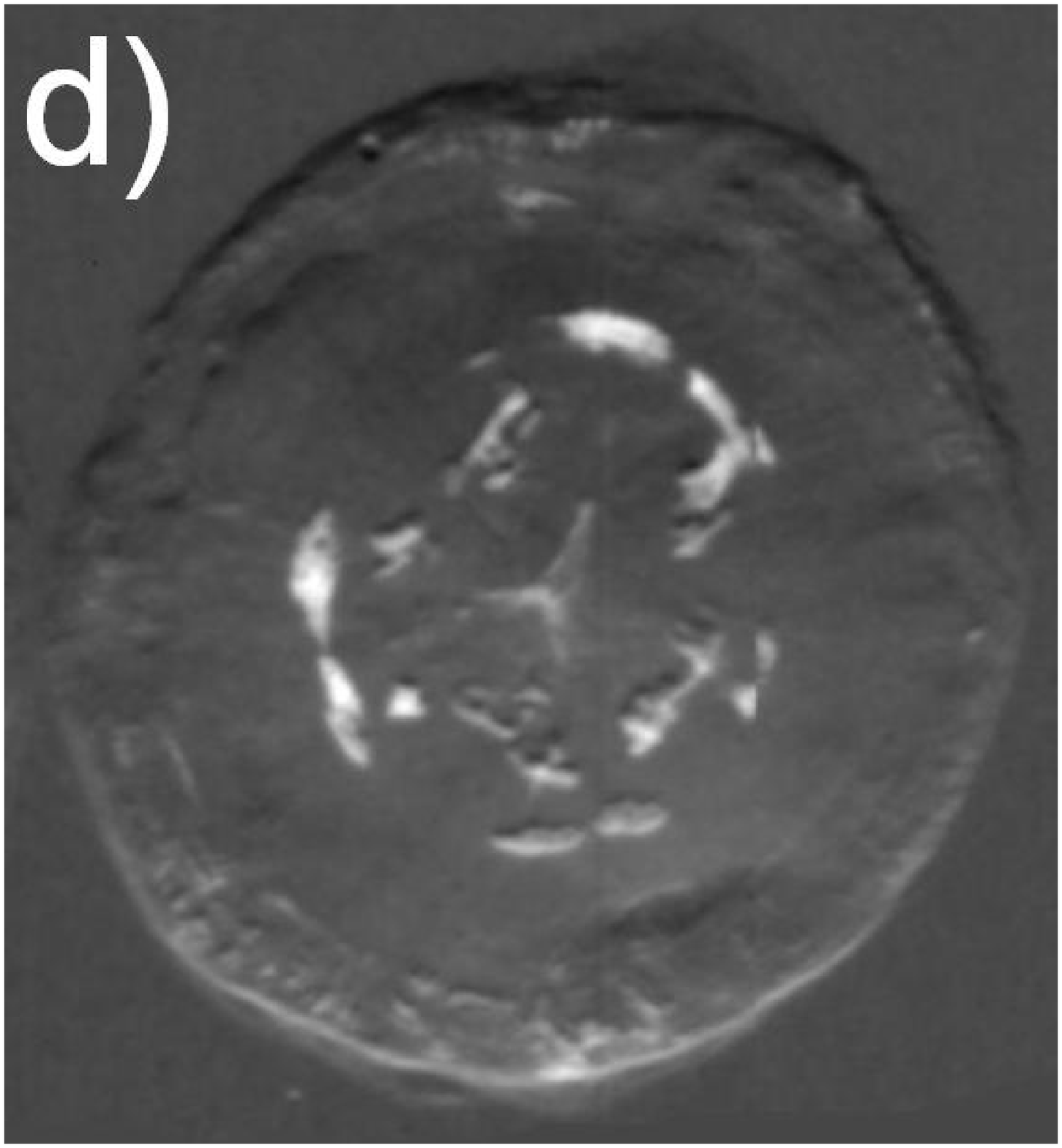}
\caption{\label{fig1} Diffraction enhanced X-ray imaging (DEI) of crystalline lenses: a) normal and b) cataractous lenses in air; and the same c) normal and d) cataractous lenses in water. Air (water) exposures reveal mostly surface (internal) structures. White areas in the cataractous lens (d) are compact aggregation of dense precipitates. X-ray photon energy is 20 KeV.}
\end{figure}

However, it is not possible to probe with visible light the internal structure of its own scattering centers and, on the other hand, electron microscopy is restricted by sample preparation procedures and small probing areas. Therefore, a detail analysis of the density fluctuations on the entire lens scale cannot be properly accomplished by optical neither electron microscopy.

X-ray absorption radiography has been applied in medical diagnosis for characterization of hard tissues such as bones and teeth. Crystalline lenses are made of soft tissues and for this reason no previous attempt of studying cataract by X-ray radiography has been reported. 

With the worldwide advances of synchrotron facilities, new X-ray imaging techniques are been developed for selected applications of interest in materials science and medicine. In one of such techniques, known as diffraction enhanced imaging (DEI), the X-ray diffraction phenomenon in crystals (Bragg diffraction) allows to separate the absorption, scattering, and refraction effects present when a monochromatic low-divergence X-ray beam is transmitted through the materials \cite{zhon2000}. This technique is very sensitive to any density fluctuation and it has already been applied for analyzing soft tissues, as for instance in cancer diagnosis \cite{chap1998}.

The present work exploits DEI for studying cataract. It also reports the finding of surprising distributions of heavy scattering centers in cataractous canine lenses. \emph{In vitro} crystalline lenses were investigated by the very short wavelength (0.062nm) of X-ray photons at 20 KeV. DEI of entire lenses are obtained at different positions of the analyzed crystal to selectively enhance the scattering, refraction, and absorption properties of the observed features. Regions of fiber cell compaction and distributions of heavy precipitates are the most common features. In some cases, dense aggregations providing absorption contrast even in the conventional radiography setup are observed. The relevance of this finding for the understanding of cataract is discussed.

\section{Results}

Significant X-ray scattering occurs at the lens surface owing to the air-lens interface. It is responsible for most of the observable features in air exposures, useful to inspect changes in the surface morphology during catarogenesis, as shown in Figs. 1(a) and 1(b) where DEI from healthy and cataractous lenses are compared.

However, to reveal internal density fluctuations the majority of the surface scattering has to be avoided. It is possible by immersing the lens in a medium of nearly the same density of the lens tissue, in this case deionized water. A very clean and uniform image is that obtained in Fig. 1(c) for water exposure of the healthy lens, while large and well defined scattering regions around the nuclei, and also in small amounts at outer regions, are observed in Fig. 1(d) for the cataractous lens.

\begin{figure}
\includegraphics[width=3.2in]{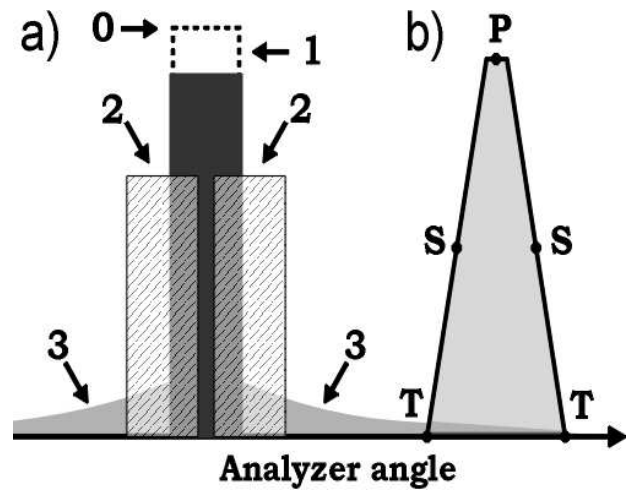}
\caption{\label{fig2} Principles of contrast in diffraction enhanced imaging (DEI) technique. a) The interaction of a monochromatic low-divergence ($\approx50 \mu$rad in the vertical direction) incident X-ray beam (0) with the tissues provides three major features on the transmitted beam: an intensity reduction (1) due to either photoabsorption and diffuse scattering; slight deviations in the beam direction (2) caused by refraction on large regions with different densities; and diffuse scattering (3) due to interface roughness and/or tiny density fluctuations as for microscopic particles. Transmitted beam images are recorded after convolution with b) the analyzer window provided by the acceptance angle for Bragg diffraction in a highly perfect crystal. Away from the analyzer tails (T) only diffuse scatterings (3) are accepted in the detection system (CCD of $50\mu$m resolution). The shoulder (S) are very sensitive to angular deviations and thus to the effect of refraction (2), while the center (C) is mostly sensitive to direct absorption (1). For X-ray photons of 20 KeV, the FWHM of the analyzer window is about 4 $\mu$rad. More details on the experimental setup can be found elsewhere. \cite{zhon2000}}
\end{figure}

As summarized in Fig.~2, the image contrast criterions of DEI allows characterization of the scattering properties of the observed internal structures in the lenses. Diffuse-scattering, refraction, and absorption images were collected respectively at the tail (T), shoulder (S), and center (C) of the analyzer window, some good examples are shown in Fig.~3. In the former type of image, the amounts of diffuse-scattering around the internal structures are consistent to the type of scattering expected for an aggregation of tiny precipitates without a well defined boundary, as in Figs. 3(a) and 3(c). On the other hand, refraction images have shown aggregations with dense cores, Fig. 1(d) and 3(d), capable to refract X-rays in an opposite sense of air bubbles, arrow in Fig. 3(f). It indicates that the precipitates are denser than the lens tissue.

\begin{figure*}
\includegraphics[width=1.6in]{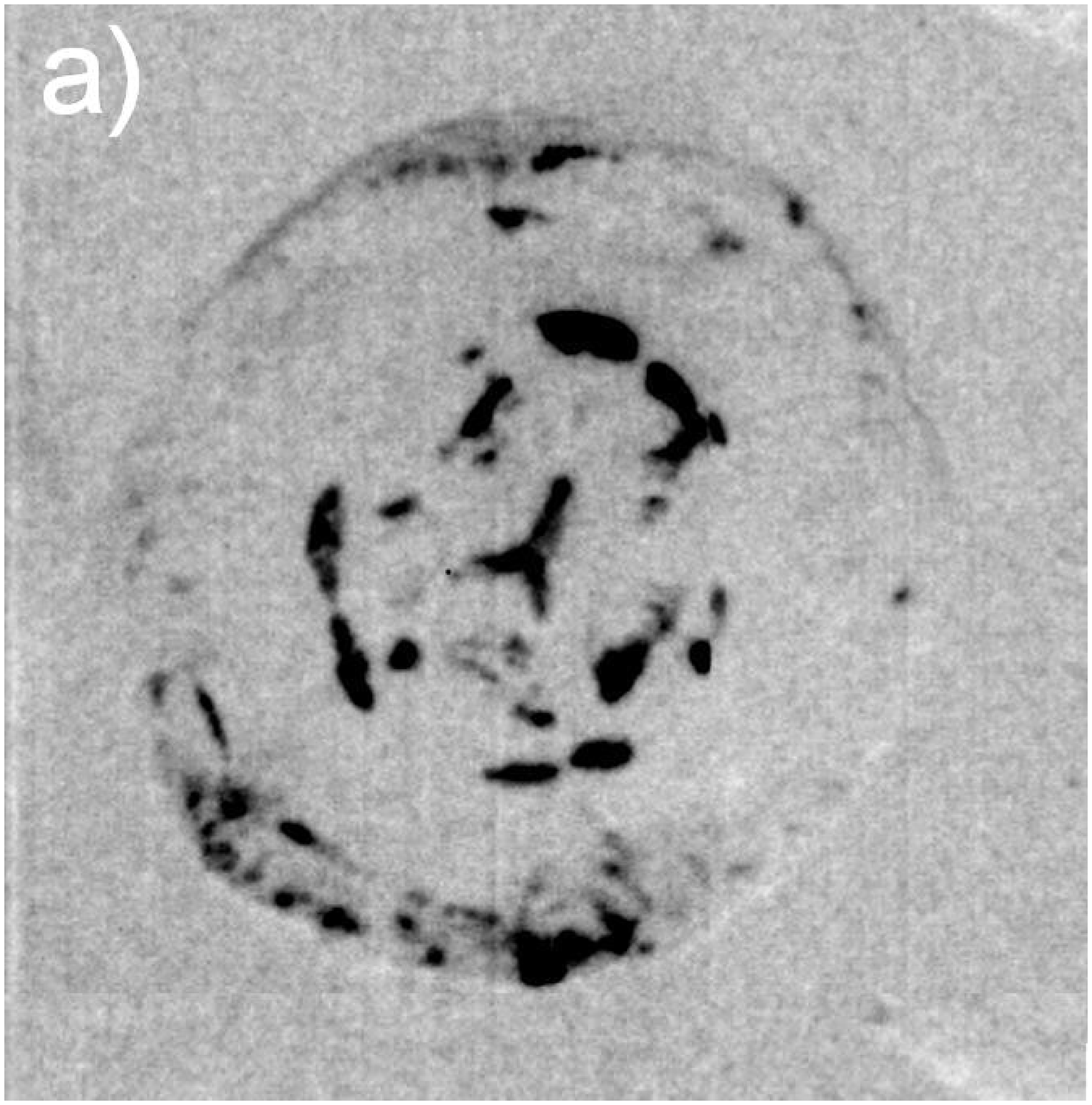}
\includegraphics[width=1.6in]{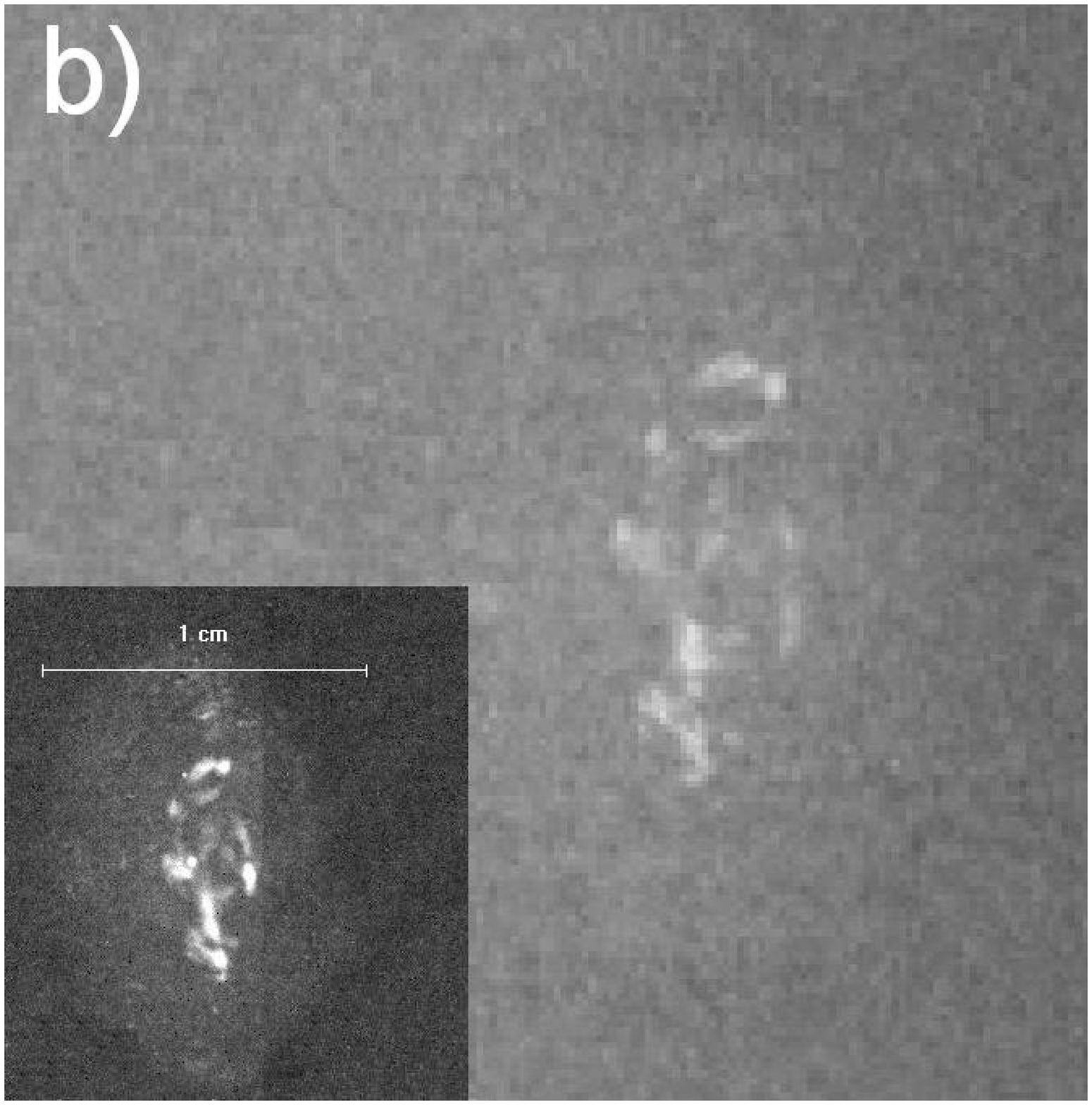} 
\includegraphics[width=1.6in]{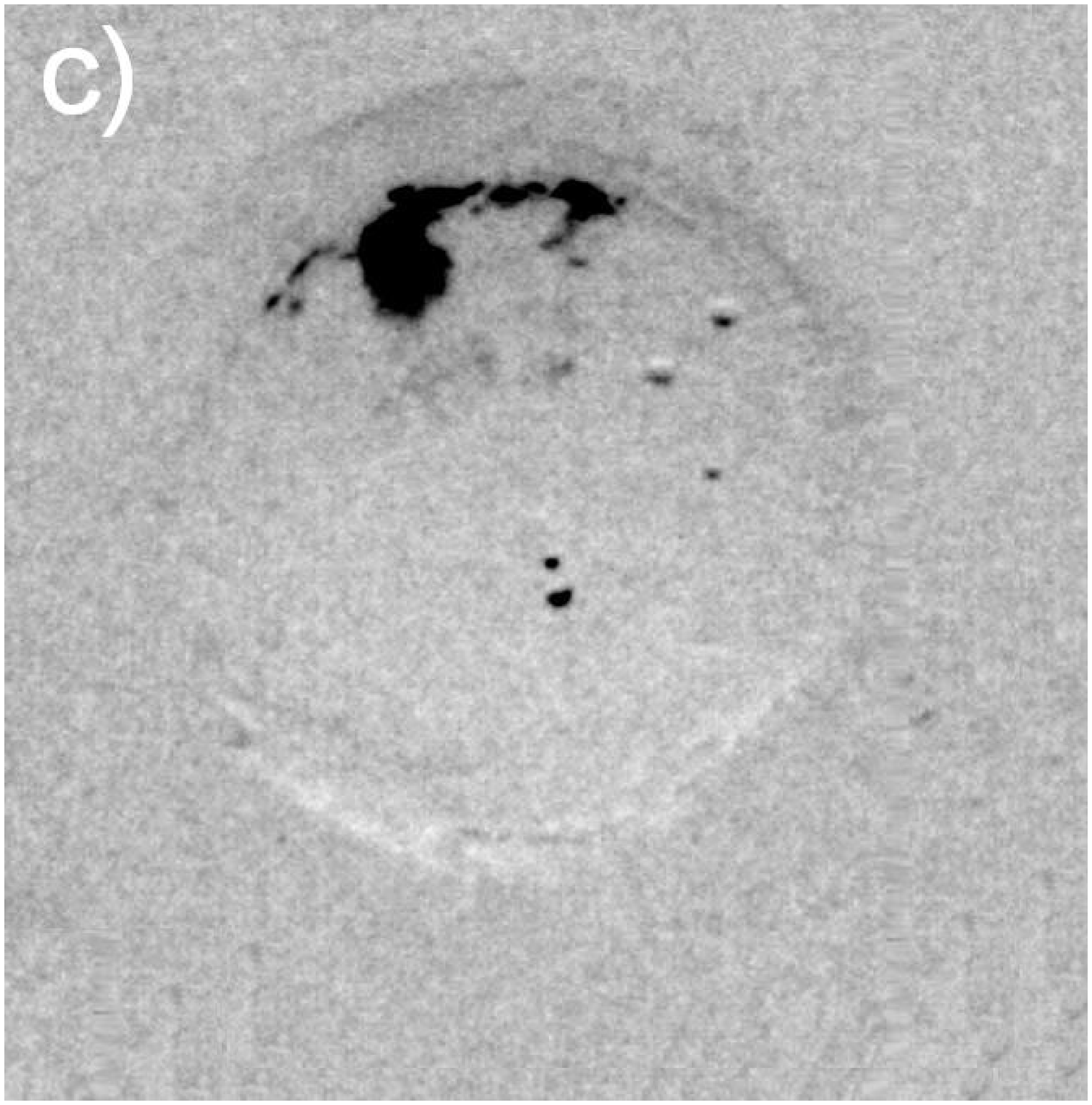}
\includegraphics[width=1.6in]{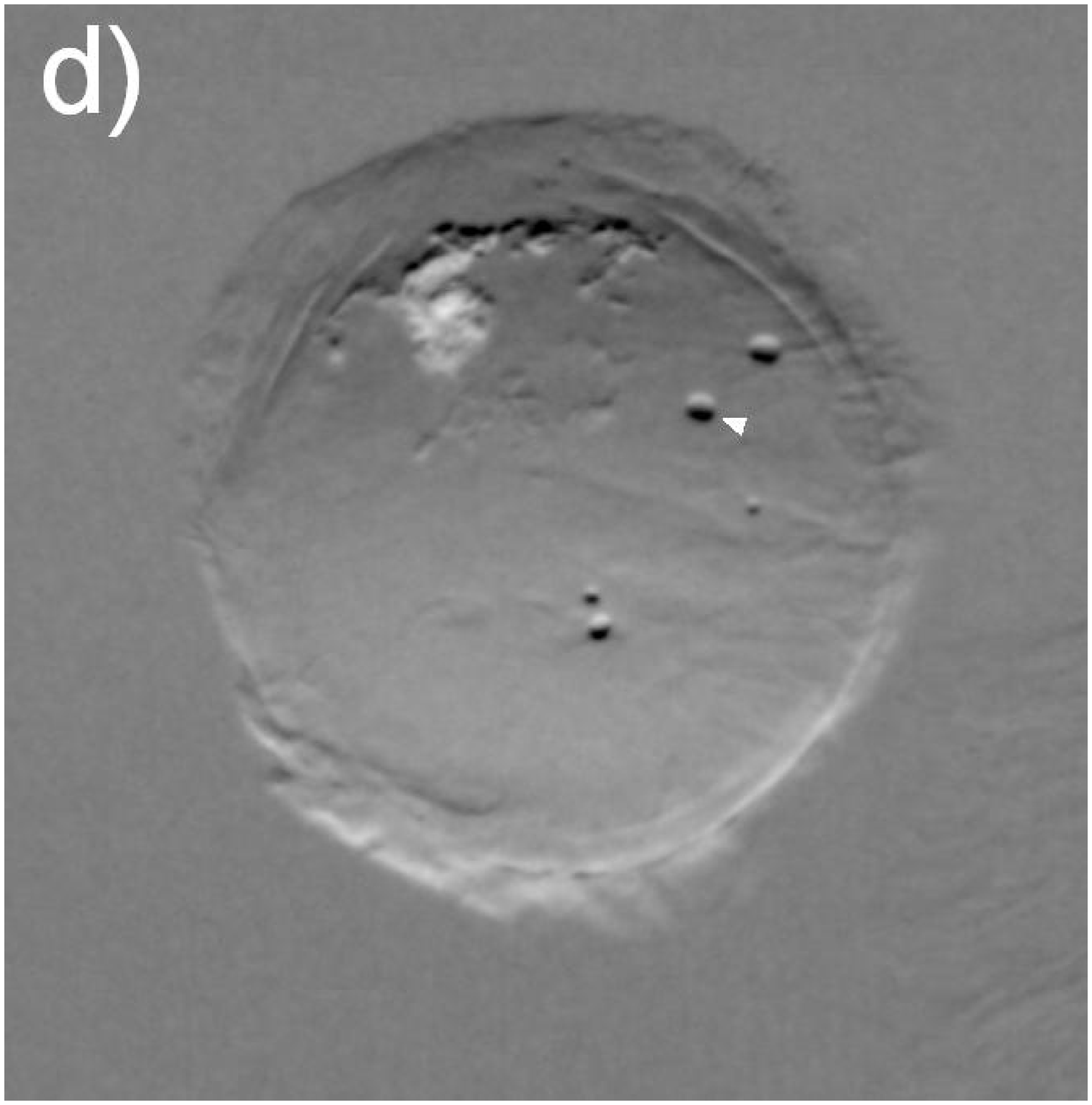} \\
\includegraphics[width=1.6in]{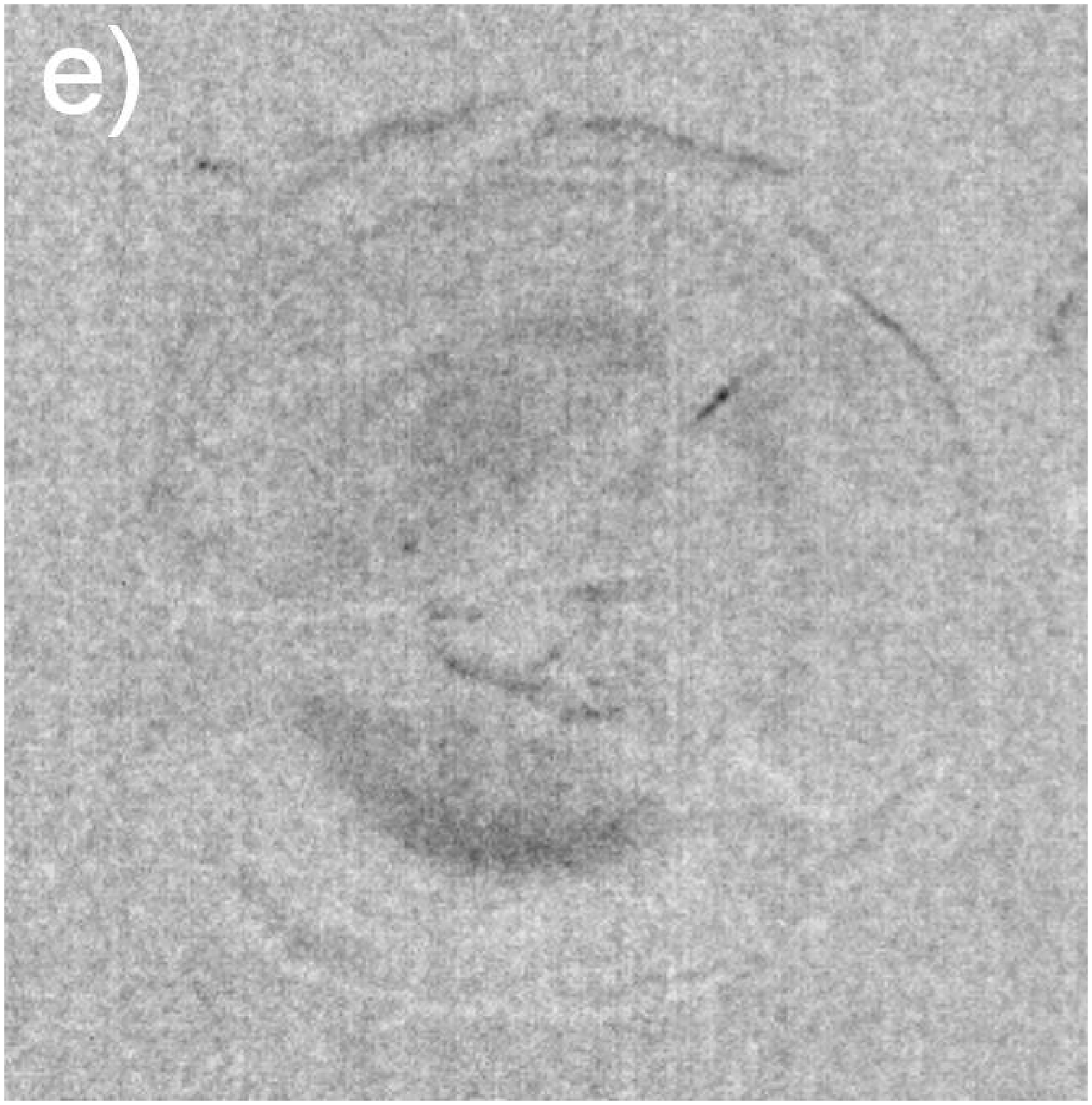}
\includegraphics[width=1.6in]{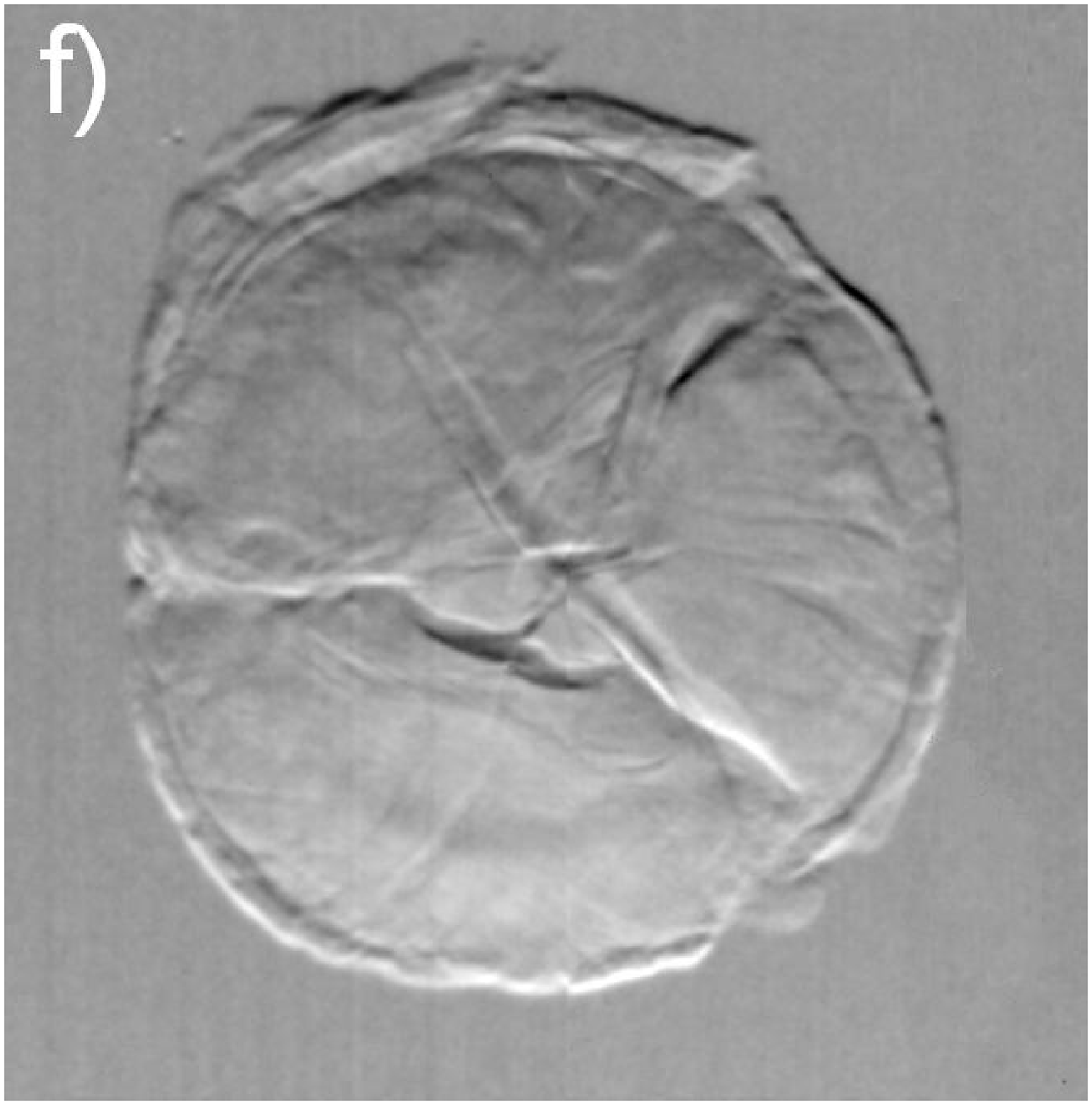} 
\includegraphics[width=1.6in]{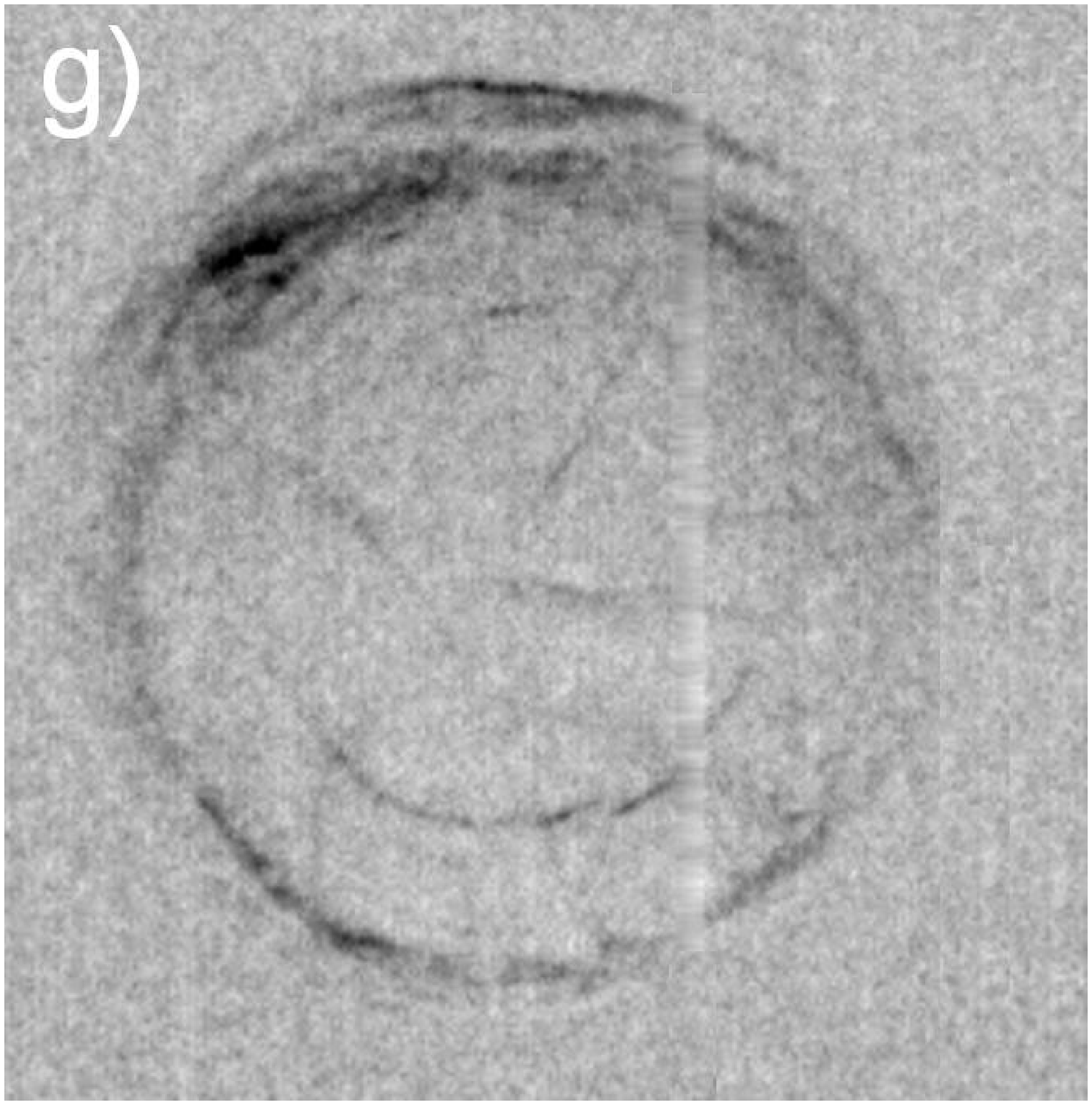}
\includegraphics[width=1.6in]{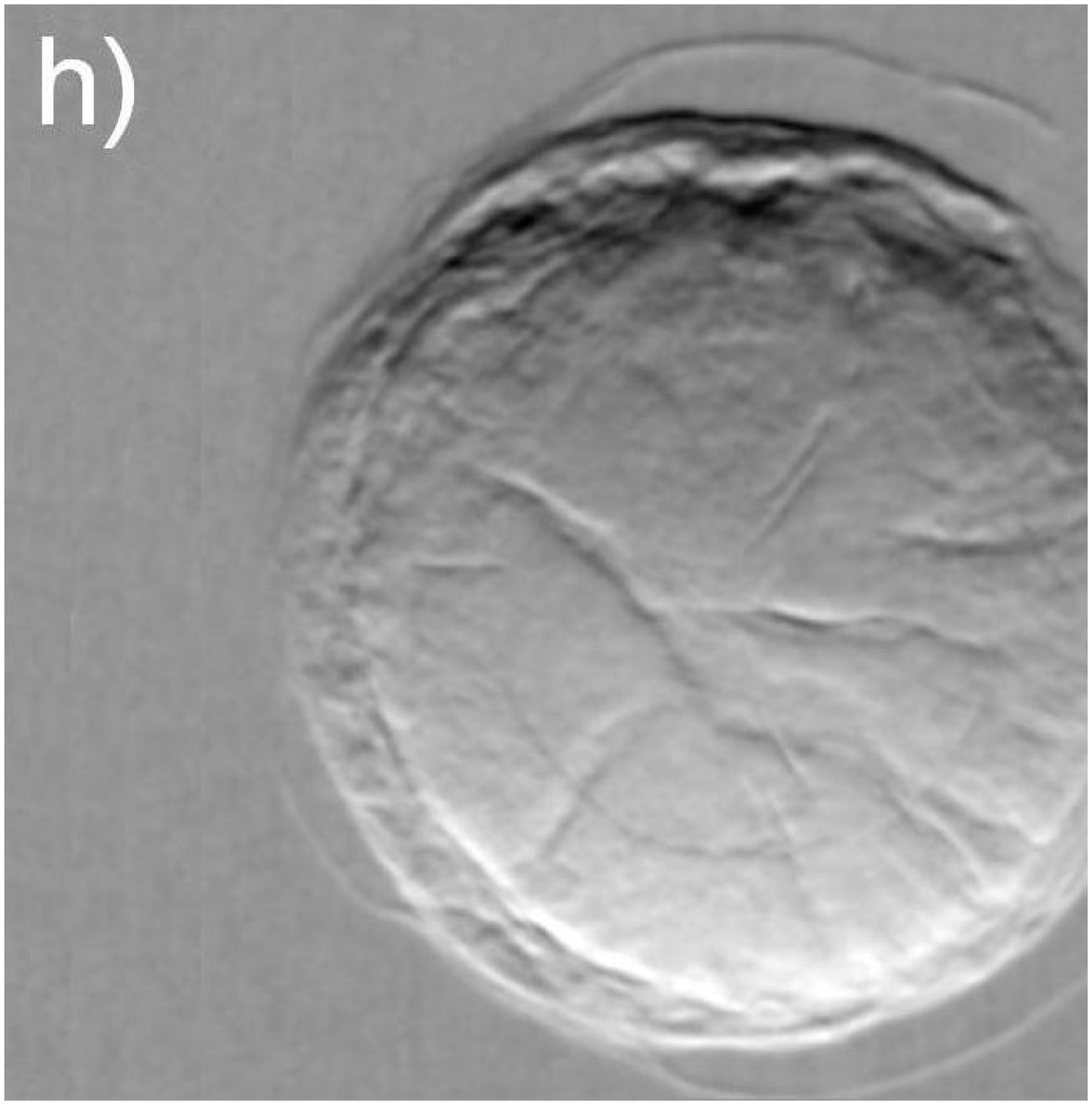} \\
\includegraphics[width=1.6in]{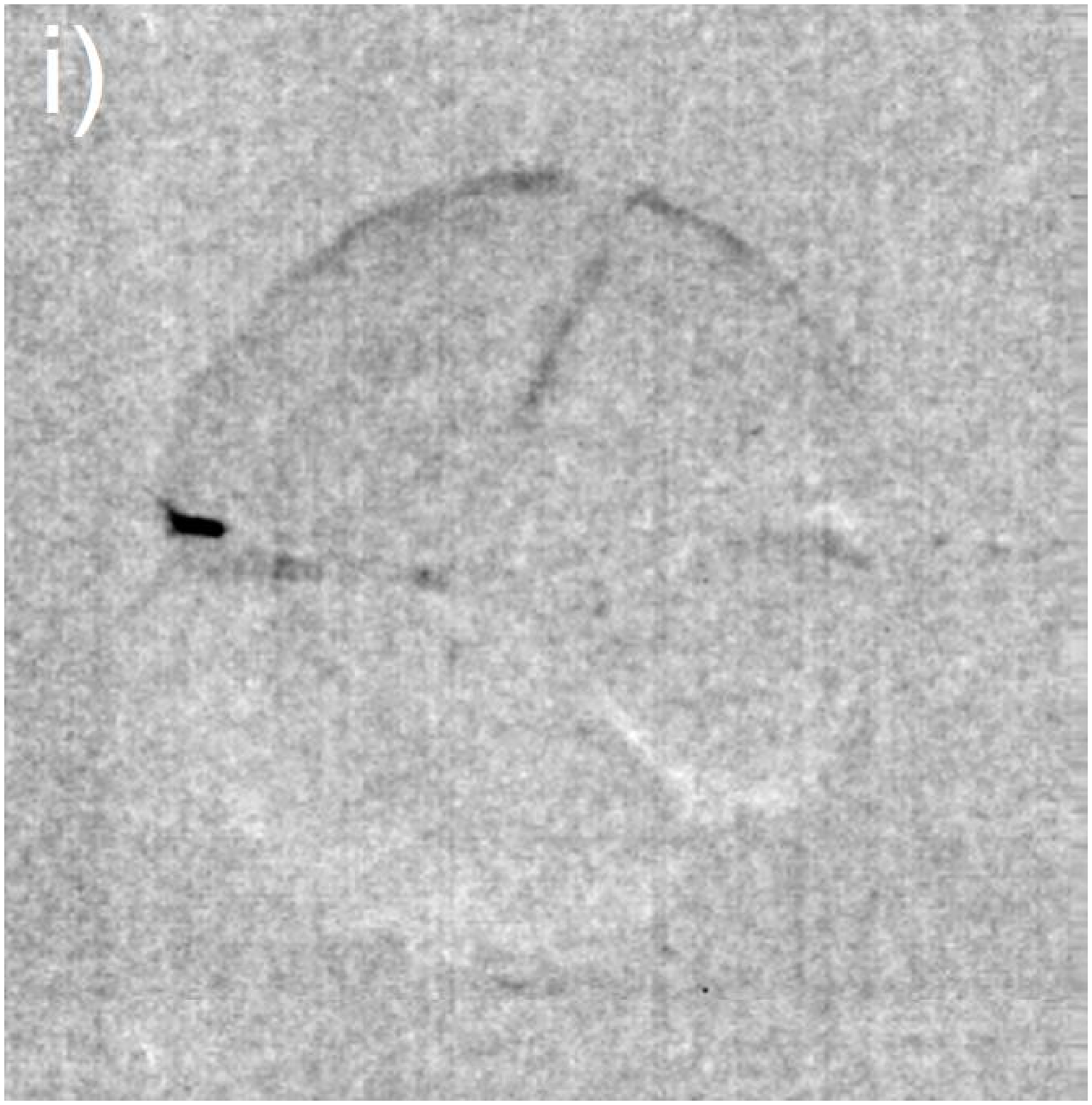}
\includegraphics[width=1.6in]{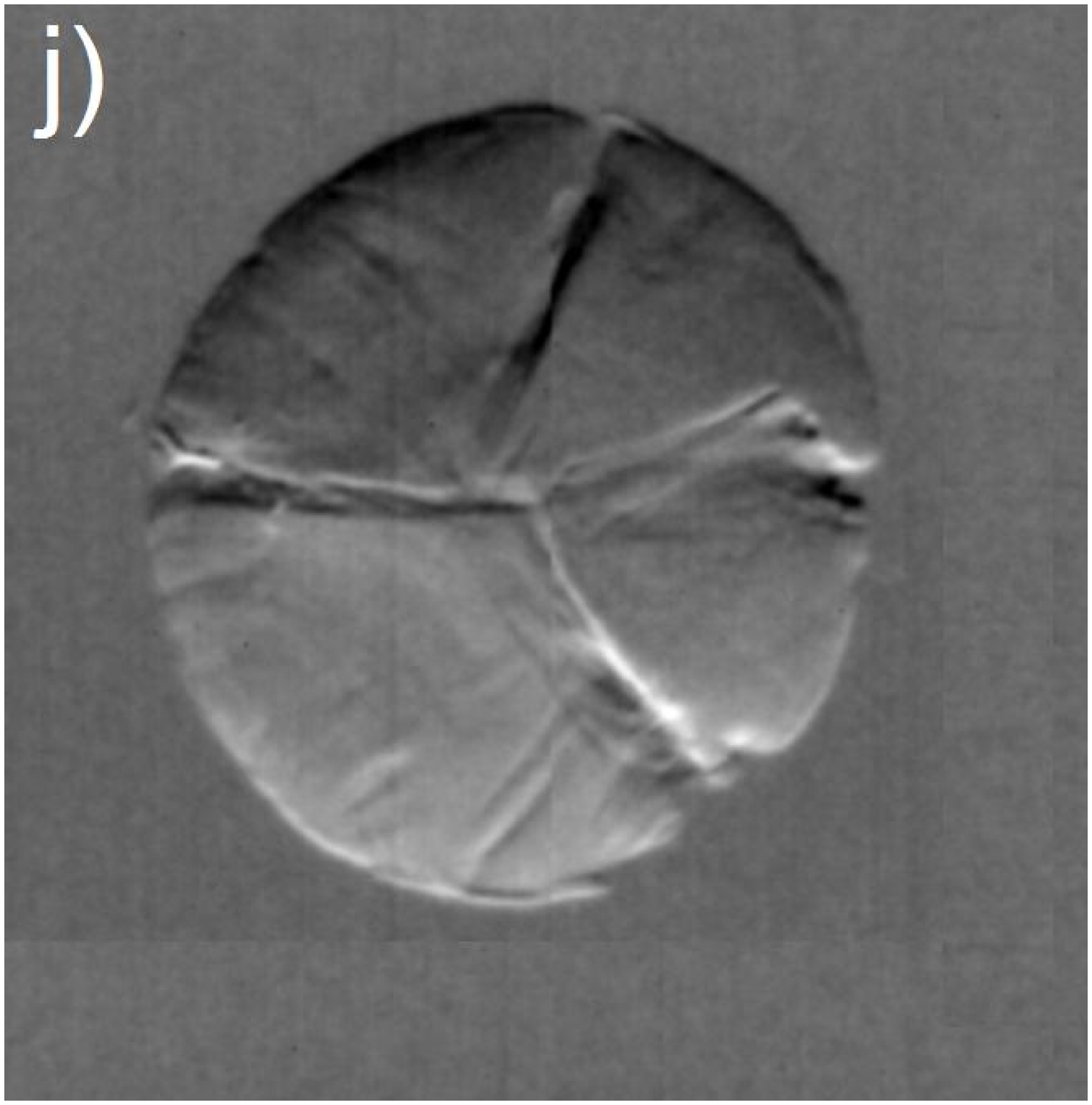}
\includegraphics[width=1.6in]{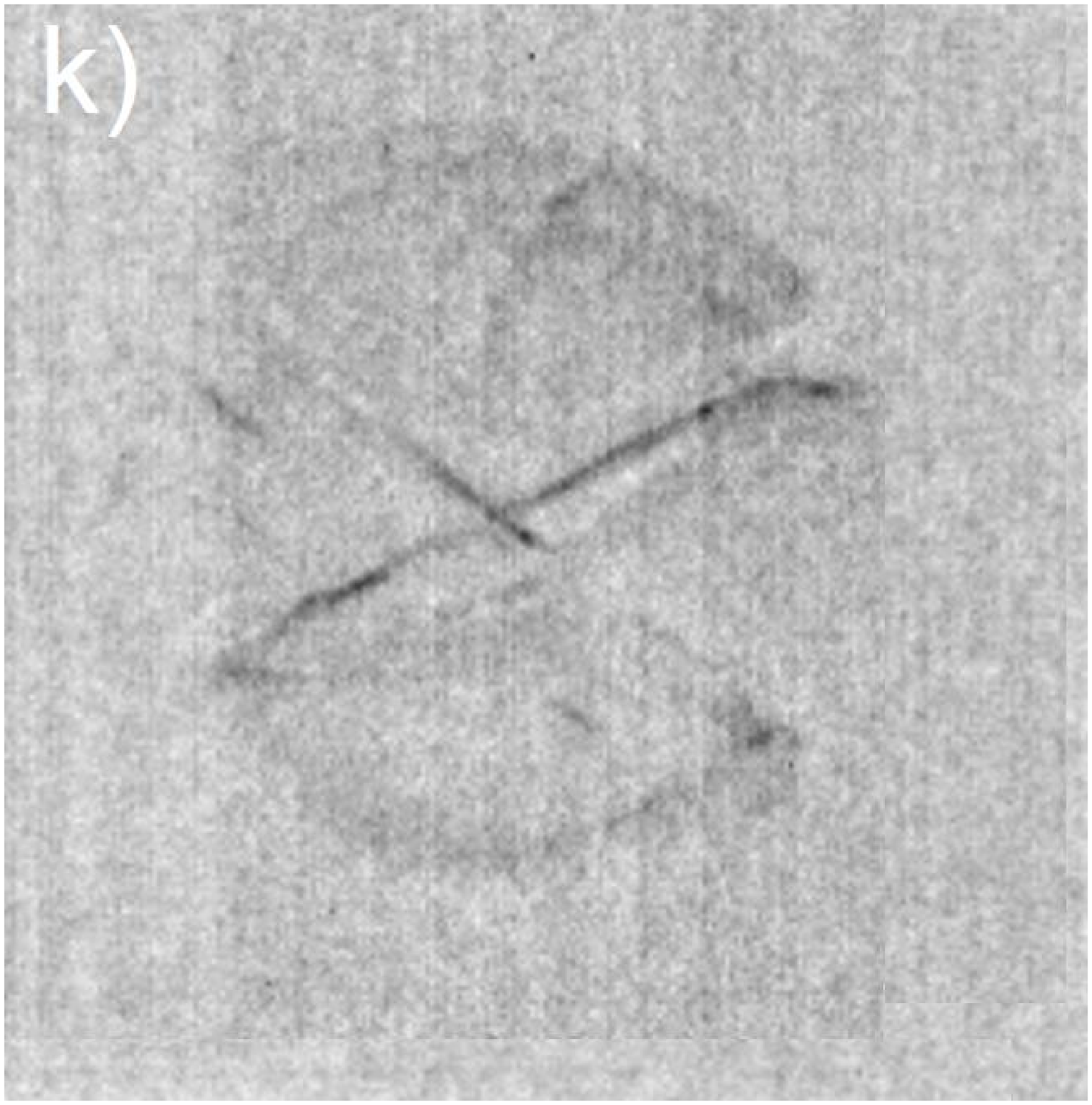}
\includegraphics[width=1.6in]{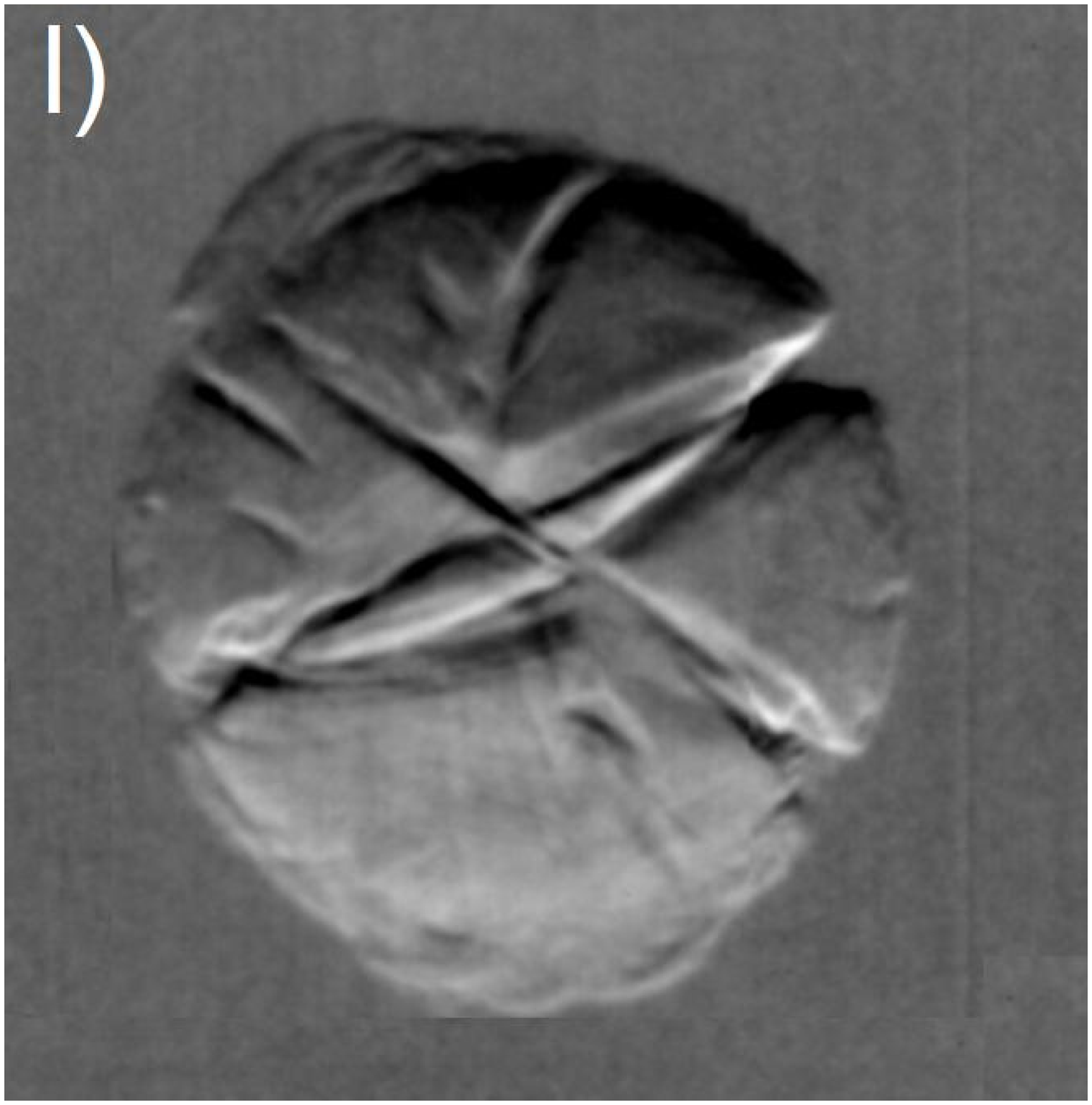}
\caption{\label{fig3} Cataractous lenses by X-ray imaging techniques. Diffuse-scattering (DS), refraction (R), and absorption (A) images corresponding respectively to the T, S, and C positions of the analyzer window in Fig.~2. a) DS-image of the same lens in Fig.~1(d), and b) side view in absorption radiographic setup (recorded on image plate); the inset shows the A-image (bar = 1 cm). c) DS-image and d) R-image of lens with well localized precipitation and dense core; the top-dark/bottom-white refraction contrast of the core is reversed in a bubble air (white arrow). e) DS-image and f) R-image of lens with dilute concentration of precipitates around the nuclei and visible Y-sutures. g) i) k) DS-images emphasizing regions of diffuse-scattering at the contour of the lenses as well as along the Y-sutures, whose marks (fiber cell compaction regions) are better visualized on h) j) l) R-images. The images were recorded on a CCD of $50 \mu$m resolution.}
\end{figure*}

By recording the transmitted beam image before the analyzer, as in conventional absorption radiography, the X-ray attenuation through aggregates with dense cores provide visible contrast in some cases. A few millimeters in length observed in frontal view, e.g. Fig.~3(a), provide a lateral attenuation in side view, Fig. 3(b), that is comparable to the attenuation of bones. Although elemental analysis is necessary to assure the composition of the precipitates, they probably are rich in calcium.

Dilute concentrations of precipitates without cores have also been observed, Figs. 3(e) and 3(f), as well as clinical cases of cataract with well-defined Y-sutures but with no distinct amounts of precipitates. In these cases, which correspond to 60\% of the analyzed cases, there are significant scattering at the suture marks that can be generated either by local compaction of fiber cells or accumulation of precipitates along the marks, as for instance in Figs. 3(g) through 3(l). Refraction images show the extension of compaction areas while diffuse-scattering images can revel the presence or not of precipitates at the sutures. Absorption images (center of the analyzer window) provide essentially the same information of diffuse-scattering ones.

Twenty clinical cases of canine cataract were investigated here, and classified by ophthalmic exam as partial or total opacity regarding the extension of cloudiness in the lens volume. All cases had severely compromised vision before surgery. No correlations to specific causes were established.

\section{Discussion}

Two distinct potential sources of light scattering in cataractous lenses have been founded by DEI. Total opacity, i.e. entire lens clouded according to ophthalmic classification, could be correlated either to visibility of Y-sutures or presence of precipitates at nuclei areas. Lenses with precipitates at areas away from the nuclei and without contrast of suture marks are the partial opacity cases.

Biochemical changes responsible for compaction of fiber cells at Y-sutures are not age-related ones. Although significant compaction occurs as a function of ageing \cite{ghou2001}, visible marks owing to compaction were observed even in cataractous lenses from young animals, Fig.~3(f). The DEI visibility of sutures and other observable features are still to be correlated to possible causes of cataract. But, independently of their causes, the visibility of sutures seems to be related to changes in the tissue at the lens scale while in the partial opacity cases the tissue changes are localized at areas with precipitates.

Mammal lenses are basically the same tissue, with very similar spectroscopic Raman signatures \cite{antu2005}, and nearly the same microstructure \cite{kusz2004}. The precipitates founded here in canine lens could also be occurring in Human lens, but it is still to be investigated by DEI. In any case, establishing the chemical composition of the precipitates as well as the mechanism responsible for their occurrence can be relevant for developing drugs capable to avoid or dissolve the formation of precipitates at early stages of partial cataract.

\section{Conclusion}

Diffraction enhanced X-ray imaging is a very powerful technique for {\em in vitro} studying distinct types of cataract. The wealthness of details available in the images can be useful for elaborating classification schemes correlating specific DEI features with potential causes of cataract such as ageing, diabetics, usage of medicines, or even ultra-violet light exposure.

\begin{acknowledgments}
The authors gratefully acknowledge Z. Zhong of the NSLS for valuable discussions and beam-time at the X15A beamline (under proposal No. 3544), and C. Parham for the expert technical assistance. This work was supported by the Brazilian agency CNPq (proc. No. 150329/2003-2 and 301617/95-3), and by the research founding offices PRP and CCInt of the Univ. of S\~ao Paulo.
\end{acknowledgments}

\end{document}